\documentclass{article}

\usepackage{arxiv}

\usepackage[utf8]{inputenc} % allow utf-8 input
\usepackage[T1]{fontenc}    % use 8-bit T1 fonts
\usepackage{hyperref}       % hyperlinks
\usepackage{url}            % simple URL typesetting
\usepackage{booktabs}       % professional-quality tables
\usepackage{amsfonts}       % blackboard math symbols
\usepackage{nicefrac}       % compact symbols for 1/2, etc.
\usepackage{microtype}      % microtypography
\usepackage{lipsum}		% Can be removed after putting your text content
\usepackage{graphicx}
\usepackage{doi}
\usepackage{siunitx}
\usepackage{placeins}

\title{The impact of frame quantization on the dynamic range of a one-bit image sensor}

\author{ \href{https://orcid.org/0000-0002-7236-7202}{\includegraphics[scale=0.06]{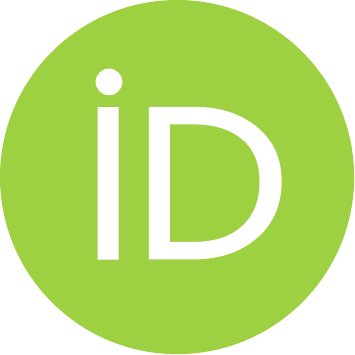}\hspace{1mm}Lucas J.~Koerner} \\%\thanks{} \\
	Department
	of Electrical and Computer Engineering\\
	University of St. Thomas\\
	St. Paul, MN 55105 \\
	\texttt{koerner.lucas@stthomas.edu} \\
}

\renewcommand{\shorttitle}{One-bit sensor frame quantization}

\hypersetup{
	pdftitle={The impact of frame quantization on the dynamic range of a one-bit image sensor},
	pdfsubject={physics.ins-det, eess.IV},
	pdfauthor={Lucas J.~Koerner},
	pdfkeywords={SPAD, Image Sensor, Dynamic Range},
}

\begin{document}
	\maketitle
	
	\begin{abstract}
		For a one-bit image sensor, the number of frames captured determines the quantization step size of the binary rate measurement. The discrete values of the binary rate set a maximum measurable intensity. In this note, we consider how this frame quantization impacts the high-end of the dynamic range of a one-bit image sensor. 
	\end{abstract}
	
	% keywords can be removed
	
\section{Background}
A single frame from a pixel of a one-bit image sensor produces a binary measurement of either 0 or 1. With zero read noise, the probability that a single frame detects a 1 (one or more photons) is given by Bernoulli statistics as $1 - e^{-H}$ and the probability of 0 is subsequently $e^{-H}$. $H$ is termed the exposure and is the the average photon count within the exposure window. The exposure relates to the photon flux ($\lambda$) and the exposure time ($T$) as $H = \lambda T$. From the average $\overline{Y}$ over a sequence of $N$ Bernoulli samples, or frames, ${Y_1, ..., Y_N}$ the exposure is estimated as $\hat{H} = -\ln(1-\overline{Y})$ \cite{maReviewQuantaImage2022}. 

\section{Maximum measurable exposure}
Since a sequence of all $1$s estimates an infinite exposure, the largest finite exposure estimate is found when one of $N$ frames measures a $0$ for a binary rate of $\overline{Y} = \hat{p} = (N-1)/N$. This maximum exposure estimate is then
\begin{equation} \label{eqn:Hqtz}
	\hat{H}_{+qntz} = -\ln \left(1- \frac{N-1}{N} \right) = \ln(N). 
\end{equation} 

The exposure referred signal to noise ratio (SNR) as a function of the binary rate $p$, exposure level $H = \lambda T$, and number of frames $N$ is given as \cite{fossumModelingPerformanceSingleBit2013, gnanasambandamExposureReferredSignaltoNoiseRatio2022}: 
\begin{equation}  \label{eqn:snr}
	\textit{SNR}(p, H, N) = - \ln(1-p) \sqrt{N} \sqrt{ \frac{1-p}{p}} = H \sqrt{N} \sqrt{ \frac{e^{-H}}{1-e^{-H}}}. 
\end{equation}
The quantization of the binary rate leads to an SNR at the high-end of the exposure found by substituting the result of Eq. \ref{eqn:Hqtz} into Eq. \ref{eqn:snr} of 
\begin{equation} \label{eqn:SNR+}
	\textit{SNR}_{qtz+} = \ln(N) \sqrt{\frac{N}{N-1}} \approx \ln(N)
\end{equation} where the right-most approximation is valid for $N\gg1$. 

One definition of Bernoulli imager dynamic range calculates the ratio of the two exposures ($H_{-}$ and $H_{+}$) that have an SNR of one \cite{chanWhatDoesOneBit2022}. To find these exposure values, Eq. \ref{eqn:snr} is rearranged and the equation  
\begin{equation} \label{eqn:H+H-}
	e^H - NH^2=1 % with SNR=1 this matches equation 26 of "What does a one-bit quanta image sensor offer"
\end{equation} is solved numerically. However, the high-end exposure value $(H_{+})$ found is not measurable as it is beyond the limit set be frame quantization given in Eq. \ref{eqn:Hqtz}.

Consider the scenario of $N=1000$ frames captured. At an $\textit{SNR}$ of $1$ $(\SI{0}{dB})$, the high-end exposure value found from Eq. \ref{eqn:H+H-} is $H_+=11.85$. However, this exposure level is highly likely to saturate. With $1000$ frames at an exposure of $H_+=11.85$ the likelihood of measuring all ones is $0.993$ as calculated from $(1-e^{-H})^N$. In other words, the high-end exposure solution ($H_{+}$) with $\textit{SNR}=1$ cannot be practically measured due to frame quantization and saturation. To measure this exposure, when frame quantization is considered, Eq. \ref{eqn:Hqtz} asserts that $140,085$ frames are required. When a frame constraint of $1000$ is maintained, Eq. \ref{eqn:Hqtz} predicts a maximum measurable exposure of $\overline{H}_{+qntz}  = \ln(N) = 6.91$ which has a $0.368$ chance of saturation.

The lower end of the exposure range for $\mathit{SNR} = 1$ approaches $H_{-} = \frac{1}{N}$ for $N \gg 1$ and $H \ll 1$. This result is found by expanding $e^H \approx H + 1$ and substituting into Eq. \ref{eqn:H+H-}. This low-end exposure is measurable as $H_{-} = \sqrt{\frac{1}{N}} = \hat{p} = 1/N$ for $N \gg 1$. 

\begin{figure}[htbp!]
	\centering
	\includegraphics[width=0.8\textwidth]{{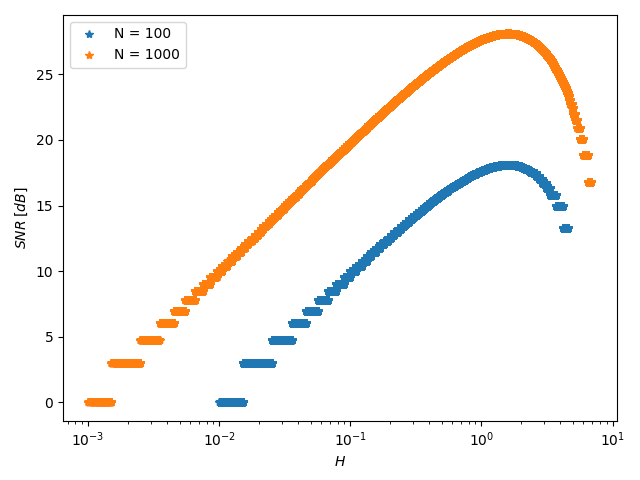}}
	\caption{The SNR of a binary imager using discrete steps of the binary rate estimate to emphasize the quantization due to the finite number of frames ($N$).} 
	\label{fig:qntzd_snr}
\end{figure}

Fig. \ref{fig:qntzd_snr} shows the exposure referred \textit{SNR} versus exposure for captures of $100$ and $1000$ frames. For a given (continuous) exposure level the representation of Fig. \ref{fig:qntzd_snr} rounds to the closest discrete binary rate ($p$) before evaluating the $\textit{SNR}$ using Eq. \ref{eqn:snr}. The curve stops at the maximum measurable exposure near $\ln(N)$ with an SNR significantly above \SI{0}{dB}.

\section{Dynamic range}

Table \ref{table:dr} shows two definitions of dynamic range. The first value is $\textit{DR} = H_{+}/H_{-}$ while the value that considers frame quantization is $\textit{DR}_{qtz} = H_{+qtz}/H_{-qtz}$. Looking across the rows of Table \ref{table:dr} shows that the dynamic range measure that considers frame quantization is consistently lower than the measure that calculates the ratio (or the logarithmic difference) of the two exposure levels with $\mathit{SNR}=1$. At low numbers of frames, the deviations between the two dynamic range measures are most significant, with around \SI{6}{dB} of difference when $100$ frames are measured. The last column shows the exposure level for a 50\% change of saturation which is evaluated as 
\begin{equation}
	H_{50\%-sat.} = -\ln \left(1 - 0.5^{1/N} \right). 
\end{equation}
An exposure limit set by a 50\% chance of saturation follows closely with $H_{+qtz}$.
\begin{table}
	\centering
	\begin{tabular}{r | rrrrrrr}
		\toprule
		N &          $H_{-}$ &  $H_{+}$ & $H_{-qtz}$ & $H_{+qtz}$ & $DR$ [dB] &  $DR_{qtz}$ [dB] & $H_{50\%-sat.}$ \\
		\midrule
		10 & \num{1.05e-01} &  5.83 & \num{1.05e-01} &  2.30 &  34.86 &  26.79 &  2.70 \\
		100 & \num{1.01e-02} &  9.00 & \num{1.01e-02} &  4.61 &  59.04 &  53.22 &  4.98 \\
		1000 & \num{1.00e-03} & 11.85 & \num{1.00e-03} &  6.91 &  81.47 &  76.78 &  7.28 \\
		10000 & \num{1.00e-04} & 14.57 & \num{1.00e-04} &  9.21 & 103.27 &  99.29 &  9.58 \\
		100000 & \num{1.00e-05} & 17.20 & \num{1.00e-05} & 11.51 & 124.71 & 121.22 & 11.88 \\
		\bottomrule
	\end{tabular}
	\caption{A comparison of two dynamic range metrics for single photon imagers. Each row assesses the minimum and maximum exposure and resulting dynamic range for a particular number of frames, $N$. The last column, \textit{$H_{50\%-sat.}$}, is the exposure value with a $50\%$ chance of being saturated (returning a one for every frame).} \label{table:dr}
\end{table}

\section{Discussion}
The compressive non-linear mapping from binary rate to intensity of one-bit image sensors expands dynamic range as compared to a linear response. This compressive transfer response suggests that one-bit sensors do not saturate. However, the intensity estimate when a one-bit image sensor measures all 1s is not defined which indicates a saturation intensity. The maximum measurable intensity level (``just before'' saturation) depends on the number of frames since the binary rate measurement is quantized in steps of $1/N$. In this note we defined the maximum measurable exposure as that estimated when the binary rate estimate is $\hat{p} = \frac{N-1}{N}$. The occurrence of saturation after a sequence of frames is probabilistic, so a precise exposure level that is guaranteed to saturate does not exist. 

Ideally, a binary imager without read noise is operated with a sufficiently rapid frame rate so that no pixels saturate. However, due to power consumption and bandwidth constraints, a practical limit on the number of frames captured may exist. The finite number of frames discretizes the measurement of the binary rate and limits the maximum measurable exposure before saturation. An open challenge in the application of binary imagers is accounting for saturated pixels and defining capture schemes that maximize performance when the number of frames is constrained.

\FloatBarrier
	
	\bibliographystyle{IEEEtran}
	\bibliography{zotero_bib.bib}  %%% Uncomment this line and comment out the ``thebibliography'' section below to use the external .bib file (using bibtex) .

	%%% Uncomment this section and comment out the \bibliography{references} line above to use inline references.
	% \begin{thebibliography}{1}
		
		% 	\bibitem{kour2014real}
		% 	George Kour and Raid Saabne.
		% 	\newblock Real-time segmentation of on-line handwritten arabic script.
		% 	\newblock In {\em Frontiers in Handwriting Recognition (ICFHR), 2014 14th
			% 			International Conference on}, pages 417--422. IEEE, 2014.
		
		% 	\bibitem{kour2014fast}
		% 	George Kour and Raid Saabne.
		% 	\newblock Fast classification of handwritten on-line arabic characters.
		% 	\newblock In {\em Soft Computing and Pattern Recognition (SoCPaR), 2014 6th
			% 			International Conference of}, pages 312--318. IEEE, 2014.
		
		% 	\bibitem{hadash2018estimate}
		% 	Guy Hadash, Einat Kermany, Boaz Carmeli, Ofer Lavi, George Kour, and Alon
		% 	Jacovi.
		% 	\newblock Estimate and replace: A novel approach to integrating deep neural
		% 	networks with existing applications.
		% 	\newblock {\em arXiv preprint arXiv:1804.09028}, 2018.
		
		% \end{thebibliography}

\end{document}